\documentclass[aps,pra,superscriptaddress,twocolumn]{revtex4-1}
\usepackage{amsmath}
\usepackage{graphicx}
\usepackage{rotating}
\usepackage{amssymb}
\usepackage{txfonts} 
\usepackage[usenames,dvipsnames]{color}

\usepackage[colorlinks=true,
            linkcolor=Blue,
            filecolor=magenta,
            citecolor=Blue,
            urlcolor=Blue,
            pdftitle={Dipolar Spins},
            pdfauthor={A. Keles},
            bookmarks=true
            ]{hyperref}

\usepackage{float}

\usepackage[numbered,open]{bookmark}

\usepackage[all]{hypcap}

\graphicspath{ {./figs/} }

\usepackage{listings}

\usepackage{tikz}

\tikzset{middlearrow/.style={
    decoration={markings,
      mark= at position 0.55 with {\arrow[scale=1,blue]{#1}} ,
    },
    postaction={decorate}
  }
}
\usetikzlibrary{decorations.pathreplacing,
  decorations.markings,decorations.pathmorphing,
    calc,
  shapes}

\usepackage{multirow}

\usepackage{pifont}

\newcommand\x{\ensuremath\mathbf{x}}

\newcommand{\dave}[1]{\langle#1\rangle_\text{\tiny$J$}}

\usepackage{bbold}

\newcommand\w{\omega}

\newcommand\rb{\mathbf{r}}

\newcommand{\oneloop}[1]
{
\begin{tikzpicture}[scale=.4,baseline=(current bounding box.center)]

    \def \x {0}
    \def \y {.1}
    \def \w {.3}
    \def \l {.5}  
    \def \dver {3}
    \draw[fill=gray] (\x,\y) rectangle (\x+\w,\y+\w);
    \draw[fill=gray] (\x+\w+\dver,\y) rectangle (\x+\dver+2*\w,\y+\w);
    \foreach \m/\n [count=\i] in {#1}
    {
      \ifthenelse{\equal{\i}{1}}
      {
        \ifx\m\n
          \draw[middlearrow={stealth}] 
        \else
          \draw[middlearrow={stealth reversed}] 
        \fi
        (\x-\l,\y+\w+\l) -- (\x,\y+\w);
        \node [left, black] at (\x-\l,\y+\w+\l) 
        {\tiny\ensuremath{\m}};
      }{}  
      \ifthenelse{\equal{\i}{2}}
      {
        \ifx\m\n
        \draw[middlearrow={stealth}] 
        \else
        \draw[middlearrow={stealth reversed}] 
        \fi
        (\x-\l,\y-\l) -- (\x,\y);
        \node [left, black] at (\x-\l,\y-\l) 
        {\tiny\ensuremath{\m}};
      }{} 
      \ifthenelse{\equal{\i}{3}}
      {
        \ifx\m\n
        \draw[middlearrow={stealth}] 
        \else
        \draw[middlearrow={stealth reversed}] 
        \fi
        (\x+2*\w+\dver,\y+\w) -- (\x+2*\w+\dver+\l,\y+\w+\l);
        \node [right, black] at (\x+2*\w+\dver+\l,\y+\w+\l) 
        {\tiny\ensuremath{\m}};
      }{} 
      \ifthenelse{\equal{\i}{4}}
      {
        \ifx\m\n
        \draw[middlearrow={stealth}] 
        \else
        \draw[middlearrow={stealth reversed}] 
        \fi
        (\x+2*\w+\dver,\y) -- (\x+2*\w+\dver+\l,\y-\l);
        \node [right, black] at (\x+2*\w+\dver+\l,\y-\l) 
        {\tiny\ensuremath{\m}};
      }{} 
      \ifthenelse{\equal{\i}{5}}
      {
        \ifx\m\n
        \draw[middlearrow={stealth reversed}] 
        \else
        \draw[middlearrow={stealth }] 
        \fi
        (\x+\w+\dver,\y+\w) 
        to [out=145,in=35]
        (\x+\w,\y+\w);
        \node [above, black] at (\x+\w+.5*\dver,\y+\w+.5)
        {\tiny\ensuremath{\m}};
      }{} 
      \ifthenelse{\equal{\i}{6}}
      {
        \ifx\m\n
        \draw[middlearrow={stealth reversed}] 
        \else
        \draw[middlearrow={stealth }] 
        \fi
        (\x+\w+\dver,\y) 
        to [out=-145,in=-35] 
        (\x+\w,\y);
        \node [below, black] at (\x+\w+.5*\dver,\y-0.5) 
        {\tiny\ensuremath{\m}};
      }{} 
    } 
  \end{tikzpicture}
}

\def\iline(#1,#2,#3,#4,#5,#6)
{
  \draw[
  thick,
  decorate,
  decoration={snake, amplitude=0.3mm, segment length=1.0mm}
  ]
  (#1,#2) -- (#3,#4); 
}
\def\fline(#1,#2,#3,#4,#5,#6)
{
  \draw[middlearrow={stealth}] (#1,#2) to (#3,#4); 
} 
\newcommand{\vertex}[1]
{
\begin{tikzpicture}[scale=.3,baseline=(current bounding box.center)]

    \def \x {0}
    \def \y {.1}
    \def \w {.5}
    \def \l {.6}  
    \def \dver {3}
    \draw[fill=gray] (\x,\y) rectangle (\x+\w,\y+\w);
    \foreach \m/\n [count=\i] in {#1}
    {
      \ifthenelse{\equal{\i}{1}}
      {
        \ifx\m\n
          \draw[middlearrow={stealth}] 
        \else
          \draw[middlearrow={stealth reversed}] 
        \fi
        (\x-\l,\y+\w+\l) -- (\x,\y+\w);
        \node [left, black] at (\x-\l,\y+\w+\l) 
        {\tiny\ensuremath{\m}};
      }{}  
      \ifthenelse{\equal{\i}{2}}
      {
        \ifx\m\n
        \draw[middlearrow={stealth}] 
        \else
        \draw[middlearrow={stealth reversed}] 
        \fi
        (\x-\l,\y-\l) -- (\x,\y);
        \node [left, black] at (\x-\l,\y-\l) 
        {\tiny\ensuremath{\m}};
      }{} 
      \ifthenelse{\equal{\i}{3}}
      {
        \ifx\m\n
        \draw[middlearrow={stealth}] 
        \else
        \draw[middlearrow={stealth reversed}] 
        \fi
        (\x+\w,\y+\w) -- 
        (\x+\w+\l,\y+\w+\l);
        \node [right, black] at (\x+\w+\l,\y+\w+\l)
        {\tiny\ensuremath{\m}};
      }{} 
      \ifthenelse{\equal{\i}{4}}
      {
        \ifx\m\n
        \draw[middlearrow={stealth }] 
        \else
        \draw[middlearrow={stealth reversed}] 
        \fi
        (\x+\w,\y) -- 
        (\x+\w+\l,\y-\l);
        \node [right, black] at (\x+\w+\l,\y-\l) 
        {\tiny\ensuremath{\m}};
      }{} 
    } 
  \end{tikzpicture}
}
\newcommand{\oneloopV}[1]
{
  \begin{tikzpicture}[scale=.8, baseline=(current bounding box.center)]
    \ifthenelse{\equal{1}{#1}}
    {
      \iline(1,0,1,1,d,a )  
      \iline(2,0,2,1,d,a )  

      \fline(0.5,0,1,0,d,a )  
      \fline(1,0,2,0,d,a )  
      \fline(2,0,2.5,0,d,a )  

      \fline(0.5,1,1,1,d,a )  
      \fline(1,1,2,1,d,a )  
      \fline(2,1,2.5,1,d,a )
    }{}  
    \ifthenelse{\equal{2}{#1}}
    {
      \iline(1.5, 0.0, 1.5, 0.3, d, a )  
      \iline(1.5, 0.7, 1.5, 1.0, d, a )  

      \fline(0.5, 0.0, 1.5, 0.0, d, a )  
      \fline(1.5, 0.0, 2.5, 0.0, d, a )  

      \fline(0.5, 1.0, 1.5, 1.0, d, a )  
      \fline(1.5, 1.0, 2.5, 1.0, d, a )  
      \draw[middlearrow={stealth}] (1.5, 0.7) to[out=0,in=0, looseness=1.5]  (1.5,0.3); 
      \draw[middlearrow={stealth}] (1.5, 0.3) to[out=180,in=180, looseness=1.5]  (1.5,0.7);
    }{}  
    \ifthenelse{\equal{3}{#1}}
    {
      \fline(1,0,1.5,.5,d,a )  
      \fline(1.5,.5,2,0,d,a )  
      \iline(1.5,.5,1.5,1,d,a )  

      \fline(0.5,0,1,0,d,a )  
      \iline(1,0,2,0,d,a )  
      \fline(2,0,2.5,0,d,a )  

      \fline(0.5, 1.0, 1.5, 1.0, d, a )  
      \fline(1.5, 1.0, 2.5, 1.0, d, a )
    }{}  
    \ifthenelse{\equal{4}{#1}}
    {
      \fline(1.0, 1.0, 1.5, 0.5, d, a )  
      \fline(1.5, 0.5, 2.0, 1.0, d, a )  
      \iline(1.5, 0.0, 1.5, 0.5, d, a )  

      \iline(1.0, 1.0, 2.0, 1.0, d, a )  
      \fline(0.5, 0.0, 1.5, 0.0, d, a )  
      \fline(1.5, 0.0, 2.5, 0.0, d, a )  

      \fline(0.5, 1.0, 1.0, 1.0, d, a )  
      \fline(2.0, 1.0, 2.5, 1.0, d, a )
    }{}  
    \ifthenelse{\equal{5}{#1}}
    {
      \iline(1,0,2,1,d,a )  
      \iline(2,0,1,1,d,a )  

      \fline(0.5,0,1,0,d,a )  
      \fline(1,0,2,0,d,a )  
      \fline(2,0,2.5,0,d,a )  

      \fline(0.5,1,1,1,d,a )  
      \fline(1,1,2,1,d,a )  
      \fline(2,1,2.5,1,d,a )
    }{}  

  \end{tikzpicture}
}

\newcommand{\siteVertexA}[1]
{
  \begin{tikzpicture}[scale=.8, baseline=(current bounding box.center)]
      \iline(1,0.25,1,.75,d,a )  

      \fline(0.5,0,1,0.25,d,a )  
      \fline(1,0.25,1.5,0,d,a )  

      \fline(0.5,1,1,.75,d,a )  
      \fline(1,0.75,1.5,1,d,a )  
    \node [above, black] at (1,.8) {\tiny\ensuremath{i_1}};
    \node [below, black] at (1,.2) {\tiny\ensuremath{i_2}};
    \node [left, black] at (0.5,1) {\tiny\ensuremath{1}};
    \node [left, black] at (0.5,0) {\tiny\ensuremath{2}};
    \node [right, black] at (1.5,1){\tiny\ensuremath{1'}};
    \node [right, black] at (1.5,0){\tiny\ensuremath{2'}};
  \end{tikzpicture}
}

\newcommand{\siteVertexB}[1]
{
  \begin{tikzpicture}[scale=.8, baseline=(current bounding box.center)]
      \iline(1,0.25,1,.75,d,a )  

      \fline(0.5,0,1,0.25,d,a )  
      \fline(1,0.25,1.5,0,d,a )  

      \fline(0.5,1,1,.75,d,a )  
      \fline(1,0.75,1.5,1,d,a )  
    \node [above, black] at (1,.8) {\tiny\ensuremath{i_1}};
    \node [below, black] at (1,.2) {\tiny\ensuremath{i_2}};
    \node [left, black] at (0.5,1) {\tiny\ensuremath{1}};
    \node [left, black] at (0.5,0) {\tiny\ensuremath{2}};
    \node [right, black] at (1.5,1){\tiny\ensuremath{2'}};
    \node [right, black] at (1.5,0){\tiny\ensuremath{1'}};
  \end{tikzpicture}
}

\begin{document}
\title{Scrambling dynamics and many-body chaos in a random dipolar spin model}
\author{Ahmet Kele\c{s}}
\affiliation{Department of Physics and Astronomy,
  University of Pittsburgh, Pittsburgh, Pennsylvania 15260, USA}
\affiliation{Department of Physics and Astronomy \& Quantum Materials Center,
  George Mason University, Fairfax, Virginia 22030, USA}
\author{Erhai Zhao}
\affiliation{Department of Physics and Astronomy \& Quantum Materials Center,
  George Mason University, Fairfax, Virginia 22030, USA}
\author{W. Vincent Liu}
\affiliation{Department of Physics and Astronomy,
  University of Pittsburgh, Pittsburgh, Pennsylvania 15260, USA}
\affiliation{Wilczek Quantum Center,
  School of Physics and Astronomy and T. D. Lee Institute,
  Shanghai Jiao Tong University, Shanghai 200240, China }
\affiliation{
         Shenzhen Institute for Quantum Science and Engineering and 
Department of Physics, Southern University of Science and Technology, 
Shenzhen 518055, China
         }
\begin{abstract}  
Is there a quantum many-body system that scrambles information as fast as a
black hole?  The Sachev-Ye-Kitaev model can saturate the conjectured bound for
chaos, but it requires random all-to-all couplings of Majorana fermions that
are hard to realize in experiments. Here we examine a quantum spin model of
randomly oriented dipoles where the spin exchange is 
governed by dipole-dipole interactions. The model is inspired by recent
experiments on dipolar spin systems of magnetic atoms, dipolar molecules, and
nitrogen-vacancy centers. We map out the phase diagram of this model by
computing the energy level statistics, spectral form factor, and
out-of-time-order correlation (OTOC) functions. 
We find a broad regime of many-body chaos where the energy levels obey
Wigner-Dyson statistics and the OTOC shows distinctive behaviors at different
times: Its early-time dynamics is characterized by an exponential growth,
while the approach to its saturated value at late times obeys a power law.
The temperature scaling of the Lyapunov exponent $\lambda_L$ shows that while
it is well below the conjectured bound $2\pi T$ at high temperatures,
$\lambda_L$ approaches the bound at low temperatures and for large number of
spins. 
\end{abstract}

\maketitle

\section{Introduction} 

Spin models with long-range interactions and
disorder have traditionally been a central playground in the study of spin glass
\cite{ed-and,PhysRevLett.35.1792,RevModPhys.58.801}.
Recently, such models have gained renewed interest thanks to the discovery of a
few remarkable properties of the Sachdev-Ye-Kitaev (SYK) model \cite{sachdev1993gapless,Kitaev-kitp}.
The Hamiltonian of the SYK model
\begin{equation}
{H}_{SYK} =  \sum_{i>j>k>l}^N J_{ijkl} \chi_i\chi_j\chi_k\chi_l
\end{equation}
describes $N$ Majorona fermions $\{\chi_i\}$ with random all-to-all couplings
$J_{ijkl}$ obeying normal distribution with zero mean and
standard deviation $\sim J/N^{3/2}$. This model can be solved exactly
in the large-$N$ limit, where it develops conformal invariance in the infrared limit
and is dual to a black hole in an emergent 1+1-dimensional spacetime \cite{Kitaev-kitp,PhysRevD.94.106002}.
The SYK model not only provides a concrete example for holographic duality in field theory,
but also sheds new light on many-body chaos and thermalization in interacting quantum systems.
For example, the model saturates the maximum bound for the onset of chaos,
conjectured to be $2\pi k_BT/\hbar$ \cite{Maldacena2016}.
Put in another way, the model scrambles information as fast as a black hole,
possibly the fastest scrambler in nature \cite{sekino2008fast}.

\begin{figure}
  \raisebox{0.5\height}
  {\includegraphics[width=0.35\columnwidth]{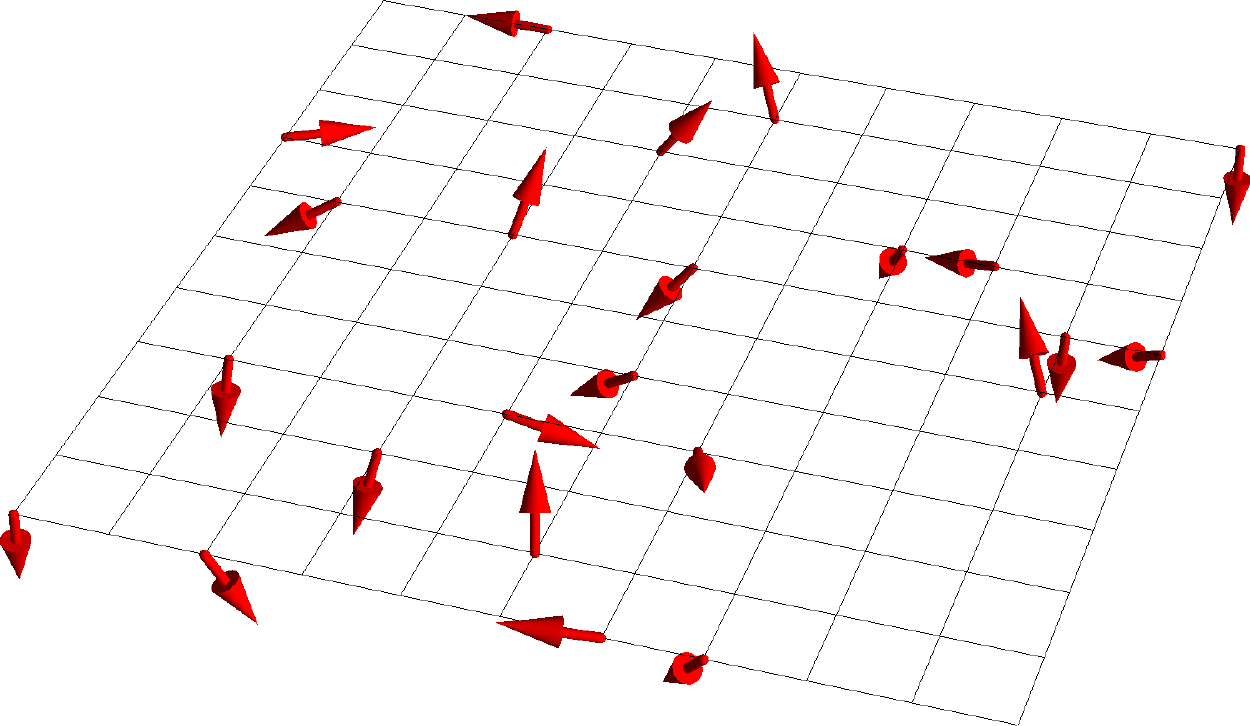}}
  \raisebox{0.5\height}
  {\includegraphics[width=0.12\columnwidth]{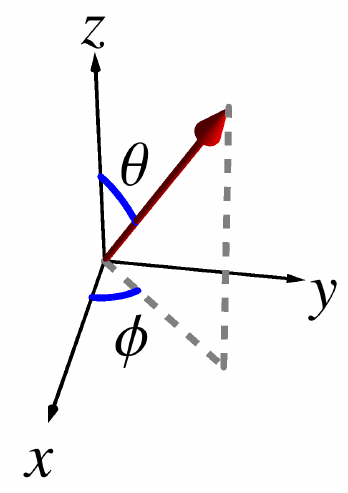}}
  \includegraphics[width=0.50\columnwidth]{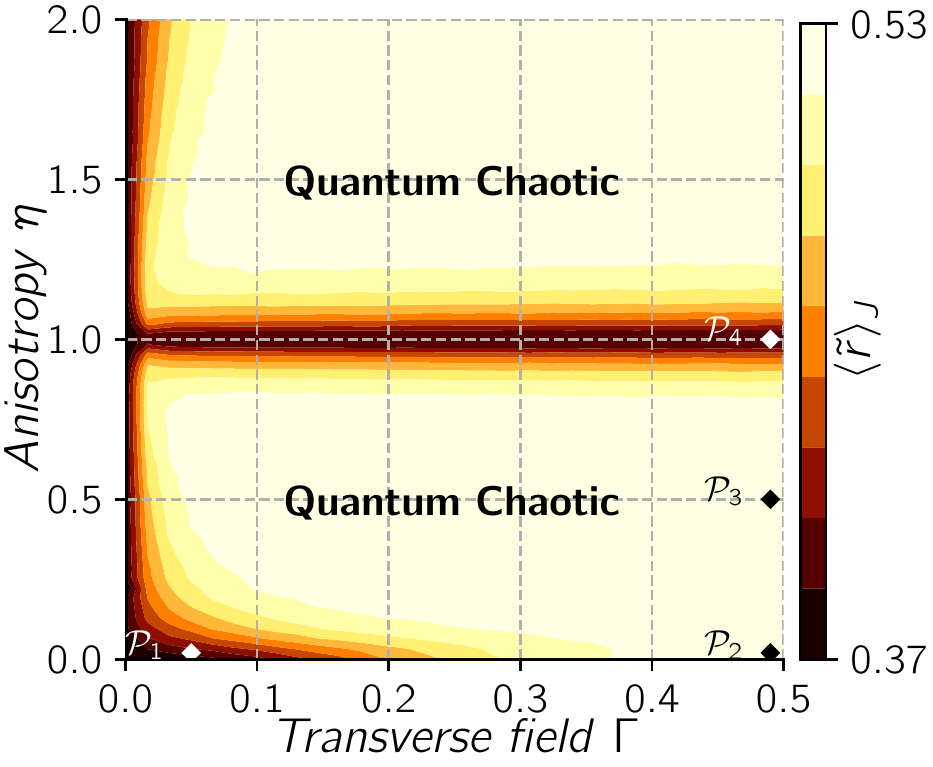}
  \caption{A random dipolar spin model on square lattice.  Left: Dipoles
    localized in randomly selected lattice sites (schematic). Each dipole
    carries spin 1/2 and has a random orientation (red arrows) parametrized by
    uniformly distributed random angles $\theta$ and $\phi$.  Right: Phase
    diagram of model in Eqs. \eqref{hd}-\eqref{dd} obtained from the energy level
    statistics of $N=10$ spins on $4\times4$ lattice averaged over 500
    disorder realization.  False color shows the disorder averaged adjacent
    gap ratio $\dave{\tilde r}$ defined in the main text. In the quantum chaotic
    (bright) region $\dave{\tilde r}=0.53$ whereas $\dave{\tilde r}=0.37$ in
    the (dark) non-chaotic region.  Four points,
    $\mathcal{P}_1$ to $\mathcal{P}_4$, are selected for detailed analysis
    which is presented below.
    }
  \label{fig:model}
\end{figure}

It remains unclear which experimental system can exhibit these intriguing properties.
Several proposals have been put forward to realize the SYK model or its variants
experimentally based on interacting fermions
\cite{Chew2017,Pikulin2017,Danshita2017,Chen2018,marino2018cavity}.
The challenge is
to engineer Majorana fermions or infinite-range coupling between complex
fermions. In this paper, we follow an alternative route and seek to find a quantum spin model
that shows fast scrambling and many-body quantum chaos similar to the SYK model.
Previous numerical study of the transverse-field Sherrington-Kirkpatrick-Ising (SKI)
model 
found exponential growth in the so-called out-of-time correlations (OTOC),
but the growth exponent $\lambda_L$, called the Lyapunov exponent in the
literature, is orders of magnitude
smaller than the maximum bound \cite{Yao2016}.
Here, we propose to relax the assumption of infinite-range coupling as in SKI and SYK, and replace it
with power-law interactions whose sign and magnitude depend on an on-site degree of freedom that
can have quenched disorder. 
This leads us to a quantum spin model inspired by
recent experiments on dipolar spin systems, e.g.
quantum gases of polar molecules \cite{yan2013observation} and magnetic atoms \cite{PhysRevLett.111.185305}
confined in optical lattices and nitrogen-vacancy centers in diamond \cite{Lukin2017}. The
dipole-dipole interaction naturally realizes long-ranged random (since
it depends on the random orientation of the dipoles) exchange between spins.

We examine the spectral statistics and OTOC of the random dipolar spin model using exact diagonalization.
In general, OTOC exhibits complicated time and temperature dependence including multiple regimes
even for idealized models such as SYK \cite{Bagrets2017} or SKI
\cite{Yao2016}. There is no single formula that can fit
all time scales or models. For example, power-law behavior
of OTOC was noted in Ref.~\cite{Chen2017} for several spin models. $\lambda_L$
for the complex fermion SYK model was found to be
very small \cite{Fu2016}. Ref.~\cite{Huizhai2017b}  reported that $\lambda_L$
reaches its maximum bound near the quantum critical points.
Key to our analysis of $\lambda_L$ is a careful separation of the exponential growth from the power
law saturation.
While our model thermalizes slower than a black hole, we find it
to be a surprisingly fast scrambler: in the limit of low temperature and large number
of spins, $\lambda_L$ seems to approach the maximum bound. 


\section{A dipolar spin model with random couplings}
Consider a generic model Hamiltonian for spin 1/2 on square lattice
\begin{equation}
  H_\mathrm{d} = \sum_{i>j} J_{ij}\biggr[ \eta (\sigma_i^x\sigma_j^x +
  \sigma_i^y\sigma_j^y)
  + \sigma_i^z\sigma_j^z \biggr]  +\Gamma\sum_i \sigma^x_i.
  \label{hd}
\end{equation}
Here $\sigma^{x,y,z}_i$ are the Pauli spin operators on site $i$,
the sum is over all occupied sites $i$ and $j$, $\eta$ is the
exchange anisotropy, and $\Gamma$ is an external field
along the $x$-direction.
In previous work, the exchange couplings $\{J_{ij}\}$ are
assumed to be random with normal distribution and all-to-all, i.e., independent of
$\rb_{ij}=\rb_i-\rb_j$, the distance between two spins at site $i$ and $j$.
We refer to model Eq. \eqref{hd} with such $\{J_{ij}\}$ as the Sherrington-Kirkpatrick XXZ model.
For $\Gamma=0$ and $\eta=1$, it reduces 
the SU(2) spin model considered in \cite{Bray1980} and generalized to SU(N) by Sachdev and Ye
\cite{sachdev1993gapless}. For $\eta=0$, it reduces to the SKI model studied in Ref.
\cite{Yao2016}.

We propose to study model Eq. \eqref{hd} for the more realistic case where $\{J_{ij}\}$ are still
random but decays as $|\rb_{ij}|^{-3}$.
To motivate such a scenario, imagine sprinkling $N$ dipoles, each carrying pseudospin 1/2,
onto the square lattice. We allow empty sites but no double occupancy.
Both the location and orientation of the sprinkled
dipoles are assumed to be random as schematically shown in Fig.~\ref{fig:model}.
These dipoles are localized with their orientations fixed.
The coupling between spins are mediated by the dipolar interaction and takes
the form
\begin{equation}
    J_{ij} = \frac{J_0}{|\rb_{ij}|^3}\biggr[\hat{d}_i\cdot\hat{d}_j
    -3(\hat{d}_i\cdot\hat r_{ij})(\hat{d}_j\cdot\hat r_{ij}) \biggr]. \label{dd}
\end{equation}
Here $\hat r_{ij}=\rb_{ij}/|\rb_{ij}|$, and the unit vector $\hat{d}_i$ specifies the orientation of the dipole at
site $i$, $\hat{d}_i = (\sin\theta_i\cos\phi_i,\sin\theta_i\sin\phi_i,\cos\theta_i)$
where angle $\theta_i$ ($\phi_i$) is random and uniformly distributed in $[0,\pi]$ ($[0,2\pi]$). The statistics of
$J_{ij}$ given by Eq. \eqref{dd} is sharply peaked at small exchanges and deviates significantly
from normal distribution as shown in Fig.~\ref{fig:stat} of Appendix~\ref{app:stats}. Roughly speaking, this is
because there is more chance to find two spins of larger distance away (and
hence small $J_{ij}$ values due to the power law decay).
The model Eqs. \eqref{hd}-\eqref{dd} is an example of strongly disordered quantum many-body systems
with long-range interaction, and our hypothesis is that it exhibits interesting scrambling dynamics.
To test this hypothesis using numerics, we consider an $L\times L$ lattice
with $N<L^2$ spins, and set the energy and inverse time units
to be $J_0$ and $\hbar=k_B=1$.

In many experiments on dipolar spin systems, a magnetic
or electric field is applied to orient all the dipole moments in the same direction $\hat{d}$ or to
set a common quantization axis for the pseudospins. This leads to
quantum spin models \cite{PhysRevB.97.245105,PhysRevLett.120.187202,PhysRevLett.119.050401} where $J_{ij}$ is simplified to $J_0 [1-(\hat{d}\cdot\hat r_{ij})^2]/|\rb_{ij}|^3$.
For randomly located dipoles $\rb_{ij}$ with vacancies present, the resultant distributions
of $J_{ij}$ is not symmetric (e.g., skewed towards negative couplings) and depends on the filling
fraction of the lattice, making the system less ideal for studying scrambling dynamics.
For this reason, we consider randomly oriented dipoles with stronger disorder described by Eq. \eqref{dd}.
This requires random local fields
to lock the dipole moments $\hat{d}_i$. Accordingly, the value of $\eta$ and $\Gamma$ may vary from
site to site. For simplicity, we assume a constant value of $\eta$ and $\Gamma$ throughout the system.


\section{Spectral statistics}
\begin{figure}[h]
  \centering
  \includegraphics[scale=0.49]{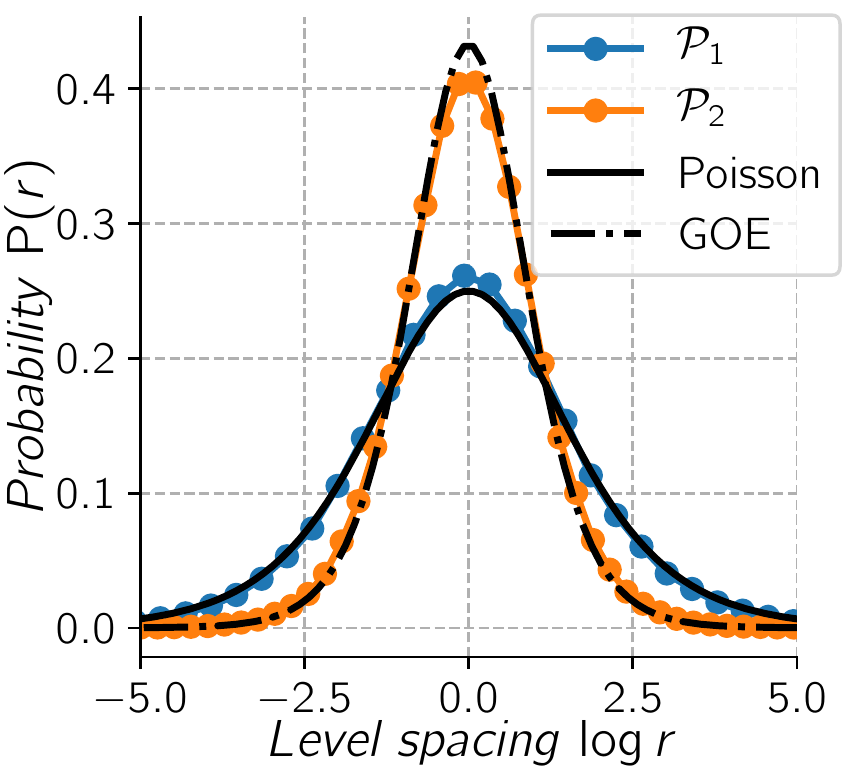}
  \includegraphics[scale=0.49]{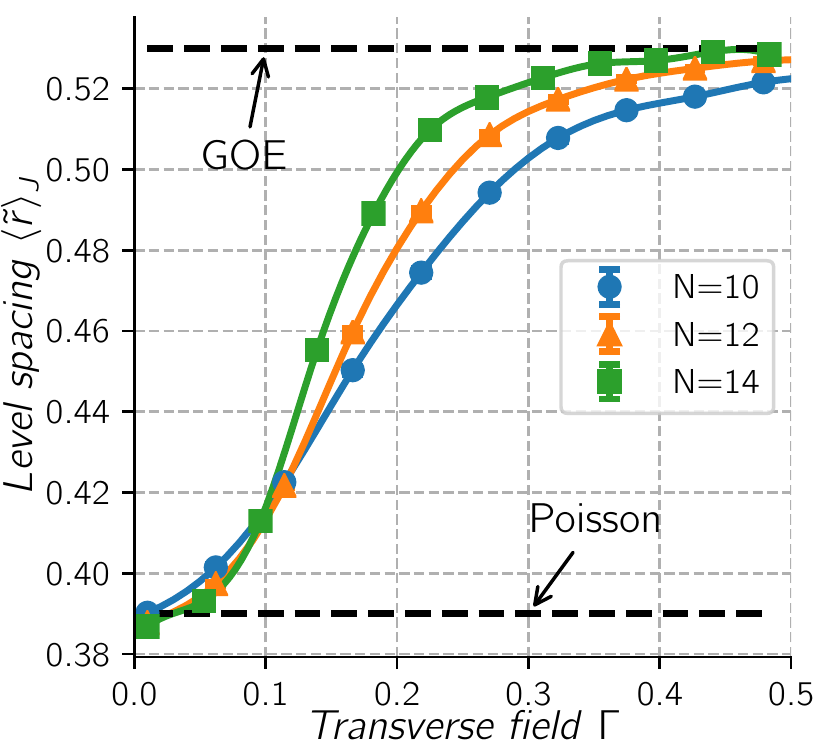}
  \caption{Energy level statistics of the random dipolar spin model in the
    Ising limit $\eta=0$. Left: Statistics of level spacing $r$ for points
    $\mathcal{P}_1$ and $\mathcal{P}_2$, calculated for 500 random realizations,
    $N=10$ and $L=4$. Also shown are the Poisson (solid line)
    and GOE (dashed line) distribution fuctions defined in the main text. For clarity, $\log r$ is used as
    the $x$-axis. Right: adjacent gap ratio $\dave{\tilde r}$ as a function of transverse field $\Gamma$ showing a transition
  to quantum chaos at critical point $\Gamma_c\sim 0.1$.  }
  \label{fig:levelstat}
\end{figure}

We first examine the energy level statistics
of this model by exactly diagonalizing $H_\mathrm{d}$ for many random
realizations of $\hat{d}_i$ for given anisotropy $\eta$ and transverse field $\Gamma$.
Then, by varying $\eta$ and $\Gamma$, we can identify different regimes (``phases") of
this model by comparing its spectra with well known behaviors of disordered
quantum many-body systems.
We adopt the spectral measures proposed in Ref. \cite{Oganesyan2007} which
are widely used in the study of many-body localization in quantum systems
\cite{Atas2013}. Let $\{E_n\}$ be the sorted energy eigenvalues
$E_1<E_2<E_3<\dots$ and $\Delta E_n=E_{n+1}-E_n$ the level spacing. Define the
ratio of adjacent level spacing $r_n=\Delta E_n/\Delta E_{n-1}$ and
$\tilde r_n = \mathrm{min}\left(r_n, r_n^{-1}\right)$.
For certain non-ergodic phases that do not thermalize, the energy levels are uncorrelated
such that the probability distribution
of $r_n$ follows Poisson distribution $\mathrm{P}(r)=1/(1+r)^2$. In comparison, for chaotic
systems the energy levels repel each other giving rise to a probability
distribution following Wigner-Dyson statistics
$\mathrm{P}(r)=Z^{-1}(r+r^2)^b/(1+r+r^2)^{1+3b/2}$ where $Z$ and $b$ are
constants depending on the ensemble symmetries. For example, $b=1$ and $Z=8/27$
for the Gaussian Orthogonal Ensemble (GOE). Note that the level statistics of the
SYK model falls into the Wigner-Dyson category and has been studied in depth \cite{You2017}.
The left panel of Fig.~\ref{fig:levelstat} shows two examples of the computed $\mathrm{P}(r)$
for our model at the same $\eta$ but different $\Gamma$. They obey Poisson and GOE
statistics respectively. Thus, this model has two distinctive phases.

The ensemble average of $\tilde{r}_n$, denoted by $\dave{\tilde r_n}$, serves as a convenient quantity to chart out
the phase diagram, since $\dave{\tilde r_n} = 0.37$ for Poisson statistics
whereas $\dave{\tilde r_n}= 0.53$ for GOE. As shown in the right panel of Fig.~\ref{fig:levelstat},
the evolution of $\dave{\tilde r_n}$ indicates a
transition from Poisson to GOE statistics for increasing $\Gamma$ in the
Ising limit $\eta=0$. Even though only a smooth crossover can be seen in numerics with finite $N$,
the variation of $\dave{\tilde r_n}$ becomes sharper as $N$ is increased.
The crossing point of the $N=10$, $12$ and $14$ curves,
 $\Gamma_c\sim 0.1$, can be taken as a rough estimate of the phase boundary for $\eta=0$ in the
thermodynamic limit $N\rightarrow \infty$. The estimated $\Gamma_c$ value is comparable to that of the
Sherrington-Kirkpatrick model \cite{Suzuki2013,Yao2016}. Similar result is
shown on a modified SYK model recently \cite{garcia2018}.
We carry out scans on the $\eta-\Gamma$ plane for $N=10$ spins averaged
over 500 random realizations. The computed $\dave{\tilde r_n }$ is illustrated
in Fig.~\ref{fig:model} (right panel) using false color where the bright
regions feature quantum many-body chaos.
We will use
independent evidences below to further corroborate this claim.
The non-chaotic regions cluster around the lines $\eta=1$ (Heisenberg limit)
and $\eta=0$ (Ising limit) where the spin model enjoys higher symmetry.
In what follows, we shall concentrate on the wide chaotic regime.

We have also computed the spectral form factor for $H_d$ which is shown in
Appendix \ref{app:form-factor}. This quantity
is defined as the disorder average $\dave{|Z(t)|^2}$, where $Z(t) = {Z^{-1}_0} \sum_n e^{-itE_n}
$ with $Z_0$ a normalization constant. It probes
spectral correlations beyond nearest level spacing, and its time evolution
shows a characteristic dip-ramp-plateau pattern for quantum chaotic systems.
For example, the SYK model shows this behavior as discussed
recently in Ref.~\cite{Cotler2017,Cotler2016}. Our results 
confirmed the presence of the dip-ramp-plateau feature in the spectral form factor
within the bright regions in 
Fig.~\ref{fig:model}. This lends additional support for identifying them
as regions of many-body chaos.


\section{Out-of-time-order correlations}
A direct diagnosis of many-body chaos 
is provided by the out-of-time-order correlations (OTOC). For two Hermitian
operators $A$ and $B$, define
\cite{larkin1969quasiclassical}
\begin{equation}
    C_{AB}(t) = -\langle[A(t),B(0) ]^2\rangle,
    \label{eq:otocAB}
\end{equation}
where in the Heisenberg picture $A(t)=U^\dagger A U$ with $U=e^{-itH}$ the time evolution operator,
and the angle bracket denotes thermal average $\langle \cdot\rangle \equiv
\mathrm{Tr}(e^{-\beta H}\cdot)/\mathrm{Tr}e^{-\beta H}$ \footnote{The definition of $C_{AB}(t)$ here
is the unregularized OTOC, which is well behaved for lattice models. For continuum models, it is better
to use the regularized OTOC, see Ref. \cite{Maldacena2016}.}.
In general,
$C_{AB}(t)$ exhibits complex dynamics at intermediate times due to the fact that
$A(t)$ and $B$ fail to commute \cite{swingle2016}. For example, the OTOC of the SYK model shows several
distinct time scales \cite{Kitaev-kitp, Maldacena2016, Bagrets2017}.
At early (pre-chaos) times up to a dissipation/collision time scale $t_d$,
the dynamics can be captured by two-point correlators 
and the growth of OTOC is negligible. This is followed by
an exponential growth $C_{AB}(t)\sim e^{\lambda_L t}$, where $\lambda_L$
is called the quantum Lyapunov exponent for convenience. Such growth is a signature of chaos.
Interestingly, the saddle point solution of SYK in the large-N limit indicates
that $\lambda_L$ saturates the conjectured bound $2\pi T$, pointing to a
possible gravity dual of this model in the form of a black hole \cite{Maldacena2016}.
Recently, it was shown that at even later times, the OTOC of SYK
crosses over to a universal power-law behavior containing $t^{-6}$
and its exact dynamics depends on temperature \cite{Bagrets2017}.

\begin{figure}[h]
  \centering
  \includegraphics[width=.49\columnwidth]{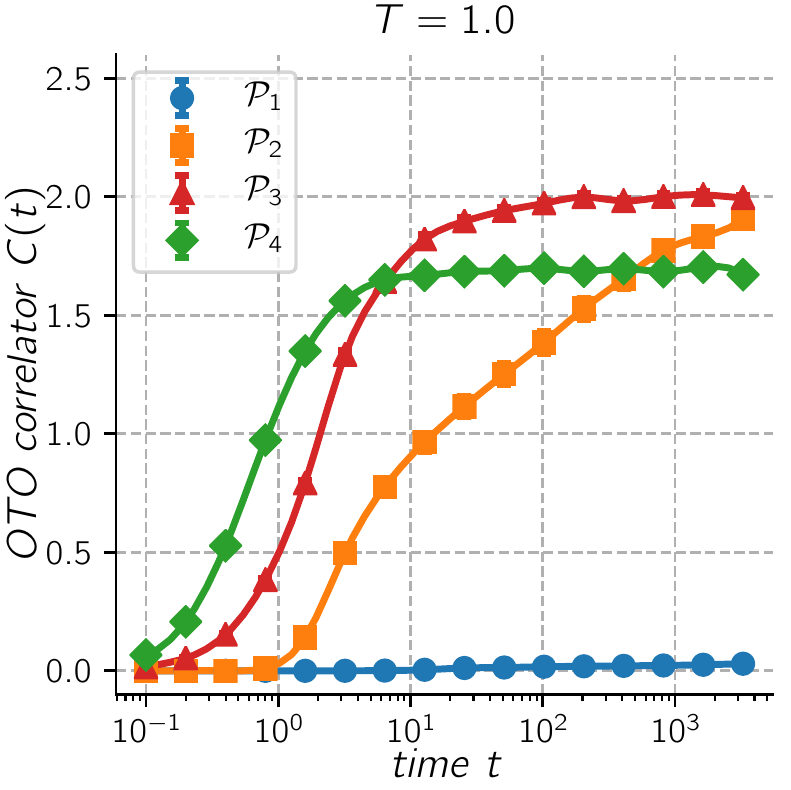}
  \includegraphics[width=.49\columnwidth]{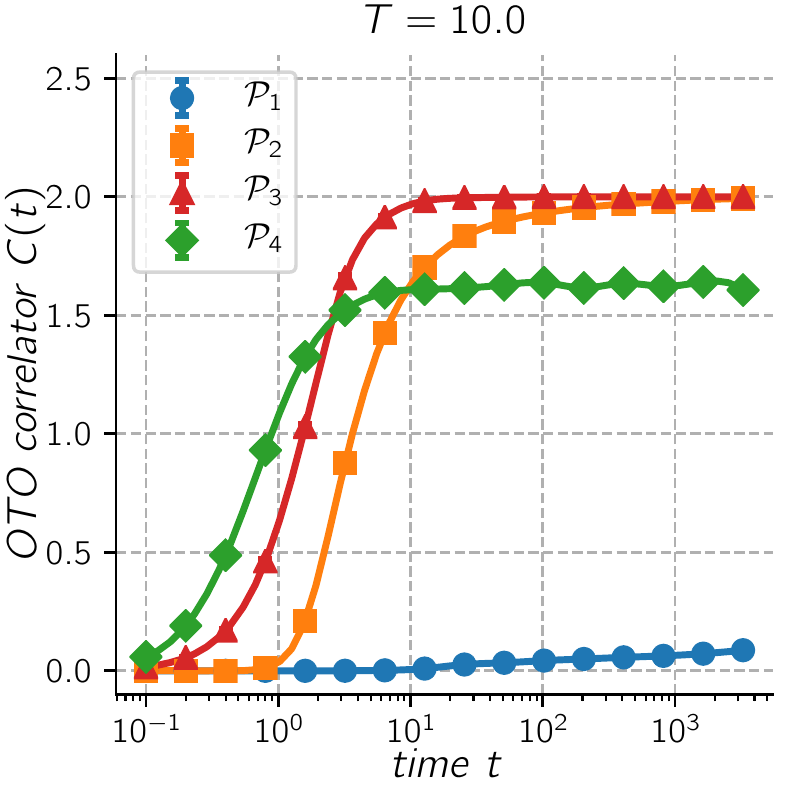}
  \includegraphics[width=.98\columnwidth]{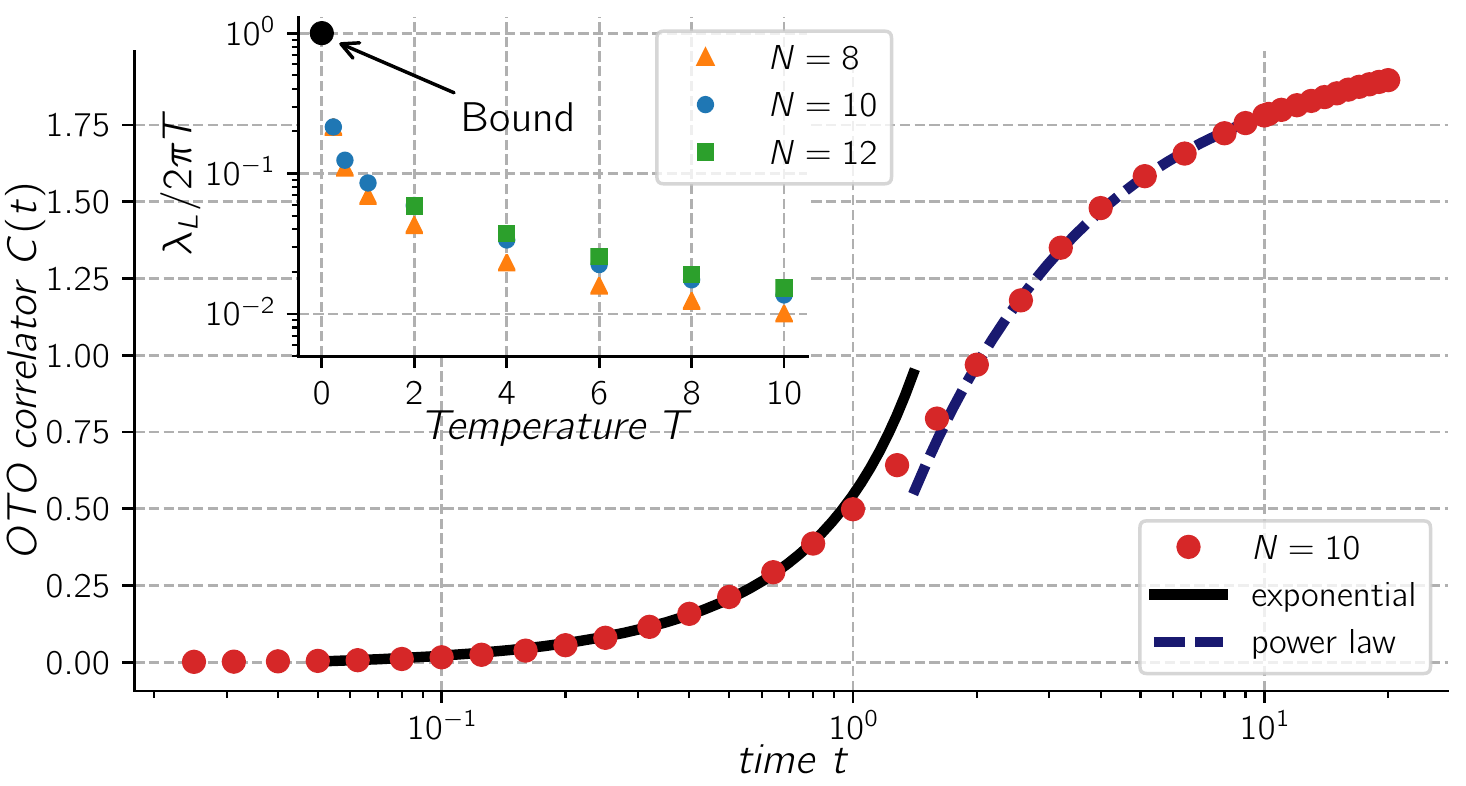}
  \caption{Out-of-time-order correlation $C(t)$ for the random dipolar spin model
  at high (top left) and low (top right) temperatures. The ensemble average
  $\dave{C(t)}$ (filled circles) 
  are obtained for $N=10$ spins and averages are calculated over 500 disorder realizations.
  Lower panel: the time dependence of $\dave{C(t)}$ for point $\mathcal{P}_3$
  and $T=1$ shows two distinct regimes. At early times, it shows exponential growth, fit
    to Eq. \eqref{exp} (black solid line) yields the Lyapunov
    exponent $\lambda_L/(2\pi T) = 0.13 \pm 0.02$. The later time dynamics obeys a power-law
    Eq. \eqref{powerlaw}, with $\mu = 0.96 \pm 0.01$ (dark blue dashed line). The inset shows the
    scaling of Lyapunov exponent with temperature for $N=8$, $N=10$ and $12$ spins. The black dot represents
    the conjectured bound for chaos, $\lambda_L/(2\pi T)=1$. The error bars
    are typically smaller than the marker sizes given in the figures.}
  \label{fig:otoc}
\end{figure}

To characterize many-body chaos in our model, we choose $A$ and $B$ as the local spin operators
$\sigma_i^z$ and $\sigma_j^z$ and define
\begin{equation}
    C(t) = \frac{1}{N(N-1)} \sum_{i\neq j}  C_{\sigma^z_i\sigma^z_j} (t),
\end{equation}
where the sum can be viewed as average over all sites.
We compute the disorder average $\dave{C(t)}$ 
where $\langle \cdot\rangle_J$ denotes
average over many random realizations. 
Fig.~\ref{fig:otoc} shows the computed OTOC for a few selected points from
the phase diagram (see Fig.~\ref{fig:model} for the location of points
$\mathcal{P}_1$ to $\mathcal{P}_4$) at two different temperatures.
For point $\mathcal{P}_1$ which has Poisson level statistics,
$\dave{C(t)}$ is very small and its growth is strongly suppressed for
both temperatures. Our result is consistent with previous observations that
the OTOC at infinite temperature grows at late times for many-body localized phases
\cite{Huang2016,Chen2017,Huizhai2017}.
For point $\mathcal{P}_3$, which is deep inside the GOE
(quantum chaotic) region, $\dave{C(t)}$ grows rapidly and saturates to
the value of 2 in the long
time limit. We will focus on this example and analyze its time and temperature
dependence in details below.
Note that the OTOC for point $\mathcal{P}_2$ resembles that of $\mathcal{P}_3$,
but the increase of $\dave{C(t)}$ begins at a later time, and
it takes longer to reach the saturation value of 2 especially at lower temperatures.
Similar results were reported for the transverse-field SKI
model~\cite{Yao2016}. While both rapid rise and saturation are observed for point $\mathcal{P}_4$, it falls
short of reaching the saturation value 2.


We now extract the Lyapunov exponent from the numerics and study its temperature
scaling, using point $\mathcal{P}_3$ deep inside the chaotic phase as
an example.
This will give a quantitative description of the scrambling dynamics, making it possible to assess
how close our model is to the ideal limit set by SYK. We find that for early times,
$t<t_L\sim 1$, the OTOC is well described by exponential growth,
\begin{equation}
C(t<t_L)=C_0 + \alpha_L e^{\lambda_L t}, \label{exp}
\end{equation}
where the fitting parameters $C_0$, $\alpha_L$ and $\lambda_L$ are obtained
from non-linear least squares fit of the data. For the $T=1$ data shown in the
lower panel Fig.~\ref{fig:otoc}, we find $\lambda_L\approx 0.13 (2\pi T)$,
below the bound $2\pi T$ as expected. Performing the same procedure for
different temperatures, we obtain the corresponding values of $\lambda_L/2\pi T$.
The result is summarized in the inset of
Fig.~\ref{fig:otoc}. Interestingly, the extracted Lyapunov exponent of our system
approaches the conjectured bound $\lambda_L/2\pi T=1$ in the low temperature
limit \footnote{Care must be taken when extrapolating toward the limit of
  $T=0$, below a critical temperature the system may enter a regime where
  chaos is absent with possible order, as demonstrated for the SYK model in
  Ref. [17].}. 

The time dependence of OTOC after $t_L$ can no longer be described by exponential.
Instead, we find that after a second time scale $t>t_P\sim 2$, the approach of OTOC to its
saturation value is well described by a power law,
\begin{equation}
C(t>t_P)=2 - \alpha_P/t^\mu, \label{powerlaw}
\end{equation}
where the parameters $\alpha_P$ and $\mu$ can be obtained from non-linear regression.
We find $\mu\approx 0.96$ for the low temperature $T=1$ data shown
in Fig.~\ref{fig:otoc}, whereas $\mu\approx 1.9$ for higher
temperature $T=10$ (fit not shown). We note these values
are quite far from the power-law exponent $\mu=6$ predicted for the SYK model in
Ref.~\cite{Bagrets2017}. Because the power law is attributed to soft mode fluctuations,
the value of $\mu$ can be model dependent.

To benchmark our numerics, we also computed the OTOC for the infinite-range Sherrington-Kirkpatrick XXZ model
and compared it with the random dipolar spin model discussed above.
The results are summarized in Fig.~\ref{fig:dipolar_vs_gaussian}  of Appendix \ref{app:comparison}. The extracted $\lambda_L$ from
both cases shows a similar temperature dependence approaching the conjectured bound $2\pi T$
at low temperatures. This offers a hint that random power-law interactions may
be
sufficient for accessing fast scrambling and the possible existence of a universal, holographic dual theory at low
energies similar to the SYK model \cite{Sachdev2015}. An interesting open
problem is to compare the growth of entanglement entropy and information scrambling
for models with different randomness, power laws, or spatial dimensions which is beyond the scope of our work. 
Such a systematic investigations using larger system sizes may reveal
phenomena beyond the SYK model.


\section{Conclusion} 
In summary, the random dipolar spin model with power-law interactions proposed and studied here
provides a concrete starting point to engineer spin systems that exhibit
rich quantum many-body dynamics. 
The phase diagram obtained from corroborating the energy level statistics, spectral form factor, and
out-of-time-order correlations points to a robust quantum chaotic phase in
this model. And the OTOC is found to show rapid
scrambling, i.e., exponential growth at early times, as well as power-law at late
times. Dipolar quantum spin systems have been realized using polar molecules such as KRb
confined in optical lattices where the pseudospin describes two rotational states of the molecules
\cite{yan2013observation,PhysRevLett.113.195302,PhysRevA.84.033619}, or
atoms with large magnetic moments where the spin refers to the
hyperfine states \cite{PhysRevLett.111.185305}. Dipolar spin models have also been achieved
using nitrogen-vacancy centers in diamond \cite{Lukin2017}, nuclear spins \cite{alvarez2015localization},
trapped ions \cite{bohnet2016quantum} with tunable interactions \cite{wilson2014tunable}, or Rydberg atoms \cite{ravets2014coherent}. 
Steady progress has been made to control the exchange coupling \cite{PhysRevB.95.024431}
and in-situ detection \cite{2017arXiv171207252C}. To realize the model proposed here, 
there remain two challenges. The first is better control of orientation randomness, possibly 
with a spatially quasi-periodic field. The second is to measure OTOC experimentally, which
has been demonstrated recently for two types
of quantum spin simulators \cite{garttner2017measuring,PhysRevX.7.031011}.
We hope our results can further stimulate experimental progress on dipolar spin quantum simulators.
\begin{acknowledgments}
We acknowledge illuminating discussions with Xiaopeng Li and Jinwu Ye.
This work is supported by 
AFOSR Grant No.  FA9550-16-1-0006 (A.K., E.Z., and W.V.L.), 
ARO Grant No. W911NF-11-1-0230 (A.K. and W.V.L.), 
MURI-ARO Grant No. W911NF-17-1-0323 (A.K. and W.V.L.), 
NSF Grant No. PHY-1707484 (A.K. and E.Z.), and
the Overseas Collaboration Program of NSF of China (No. 11429402) sponsored by
Peking University (W.V.L.).
The numerical calculations are carried out on the ARGO clusters provided by the
Office of Research Computing at George Mason University and supported in part by NVIDIA Corporation.
\end{acknowledgments}
\appendix
\section{Statistics of exchange couplings}
\label{app:stats} 

In the SYK model, the statistics of four
fermion couplings $\{J_{ijkl}\}$ obeys normal distribution, they have zero mean and
a constant standard deviation. Thanks to this, one can apply the replica trick and
perform integration over couplings which yields an exact solution. The exchange
couplings in infinite-range (e.g. Sherrington-Kirkpatrick) spin models studied so far also obey normal distribution.
Application of the replica trick suggested a possible spin
glass ground state \cite{Bray1980}. Ref.~\onlinecite{sachdev1993gapless} further generalize the
SU(2) spin model to SU(N)
spins and found a saddle point with disordered (spin liquid) state for large $N$.

\begin{figure}[h]
  \centering
  \includegraphics[width=0.3\textwidth]{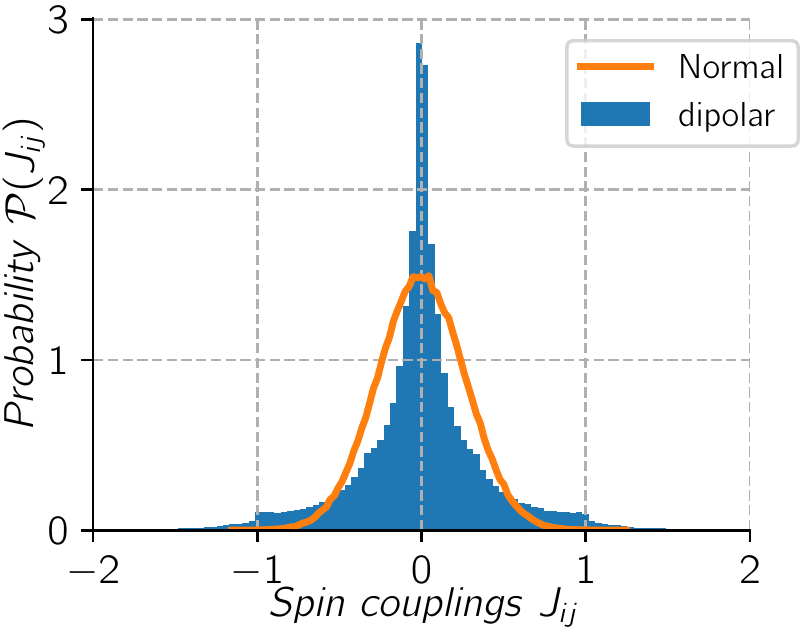}
  \caption{Statistics of dipolar spin exchange $J_{ij}$ generated from a few
    thousand random samples. For each sample, we consider $N=12$ dipolar spins
    localized in $4\times 4$ square lattice with random positions as well as
    random dipolar orientations. It deviates greatly from the normal
    distribution. }
  \label{fig:stat}
\end{figure}

We show the statistics of
couplings $J_{ij}$ 
in Fig.~\ref{fig:stat} 
for the spin-1/2 model $H_d$ introduced in the main text based on randomly
oriented dipolar spins. The histogram significantly deviates from Gaussian distributions
(the orange curve is a Gaussian constructed using the mean and standard deviation from the data).
It follows more closely the Cauchy-Lorentz distribution. This suggests that the replica trick
may not be applied straightforwardly to $H_d$. To enhance quantum fluctuations and
prevent a possible spin glass ground state, we include a transverse field and keep
$\Gamma$ as a tuning parameter in our model.

\section{Spectral Form Factor}
\label{app:form-factor} 

Spectral form factor is defined as the thermal expectation value of time evolution operator
$U=e^{-i t H}$,
\begin{equation}
    Z_\beta(t) = \mathrm{Tr}\left(e^{-\beta H - i t H}\right)/
    \mathrm{Tr}(e^{-\beta H}),
\end{equation}
where $H$ is the Hamiltonian under consideration, $t$ is time, and
$\beta$ is the inverse temperature. Here we are interested in the
infinite temperature limit, so we define
$Z=Z_{\beta\rightarrow 0}$. In the energy eigen-basis, one can expand
the trace to get
\begin{equation}
    Z(t) = \frac{1}{Z_0} \sum_n e^{-itE_n}
\end{equation}
where $Z_0$ is the normalization constant, and $E_n$ are the energy eigenvalues. At early times, $Z(t)$
approaches unity whereas at late times the phase factor inside the sum rapidly fluctuates.
When ensemble average is taken, $Z(t)$ vanishes for large $t$. Therefore, it is more useful to study
the modulus square of $Z(t)$,
\begin{equation}
    |Z(t)|^2 = \frac{1}{Z_0^2} \sum_{n,m} e^{-it(E_n-E_m)}.
\end{equation}
Notice that this quantity contains information about the level spacing $E_n-E_m$ which may obey
non-trivial statistics such as Wigner-Dyson distribution. In addition, it also probes
level correlations beyond the nearest neighbors in the energy level spectrum. This offers certain
advantages over the level spacing measures $P(r)$ and $\tilde{r}_n$ discussed in the main text.
\begin{figure}[h]
  \centering
  \includegraphics[width=0.23\textwidth]{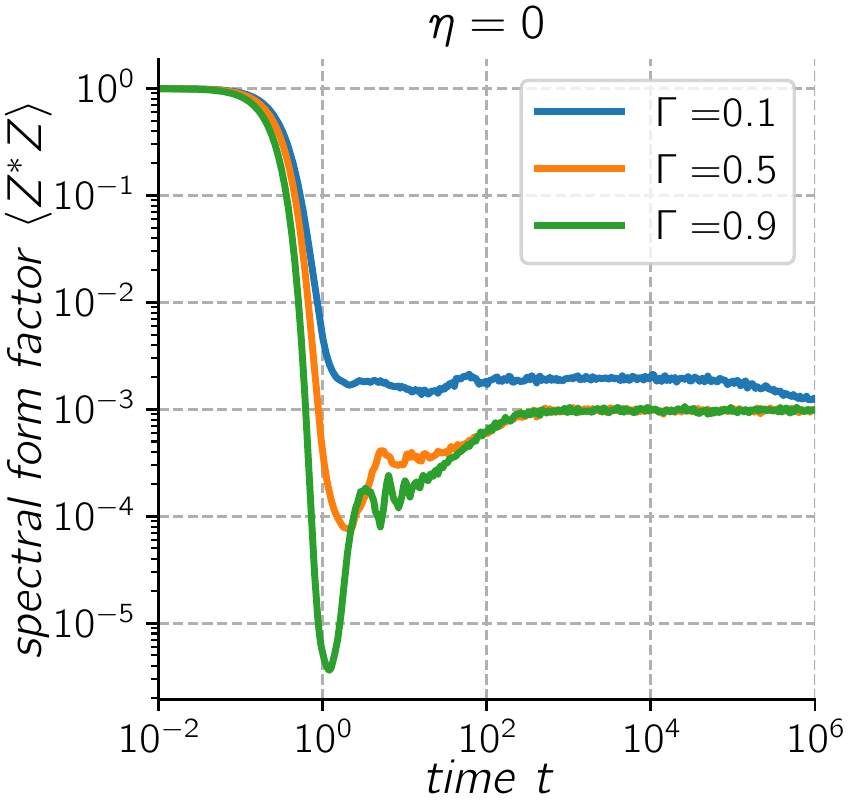}
  \includegraphics[width=0.23\textwidth]{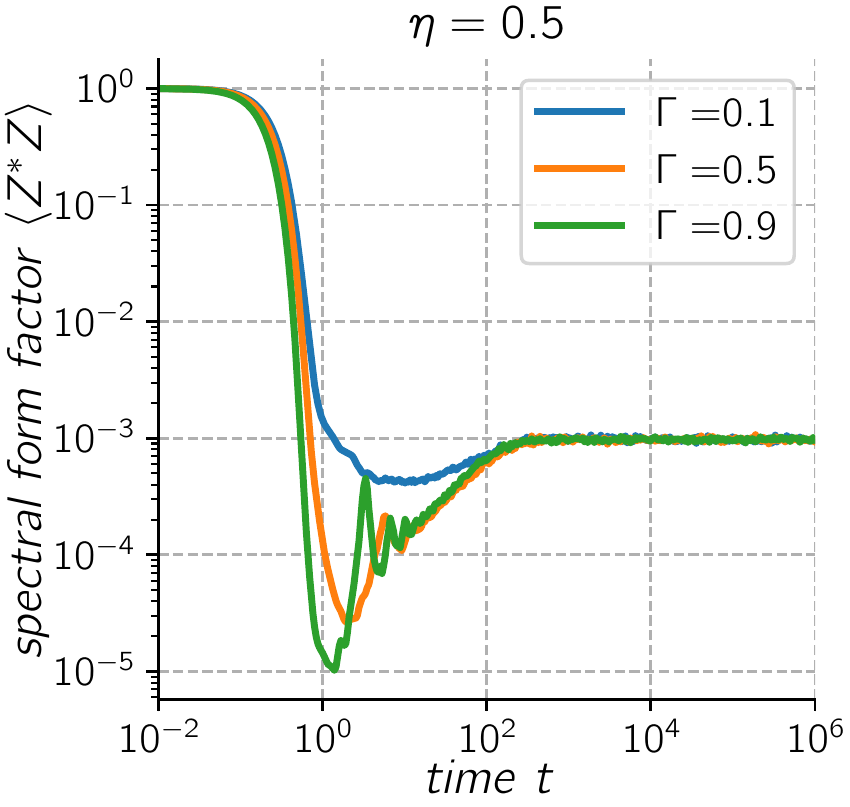}
  \includegraphics[width=0.23\textwidth]{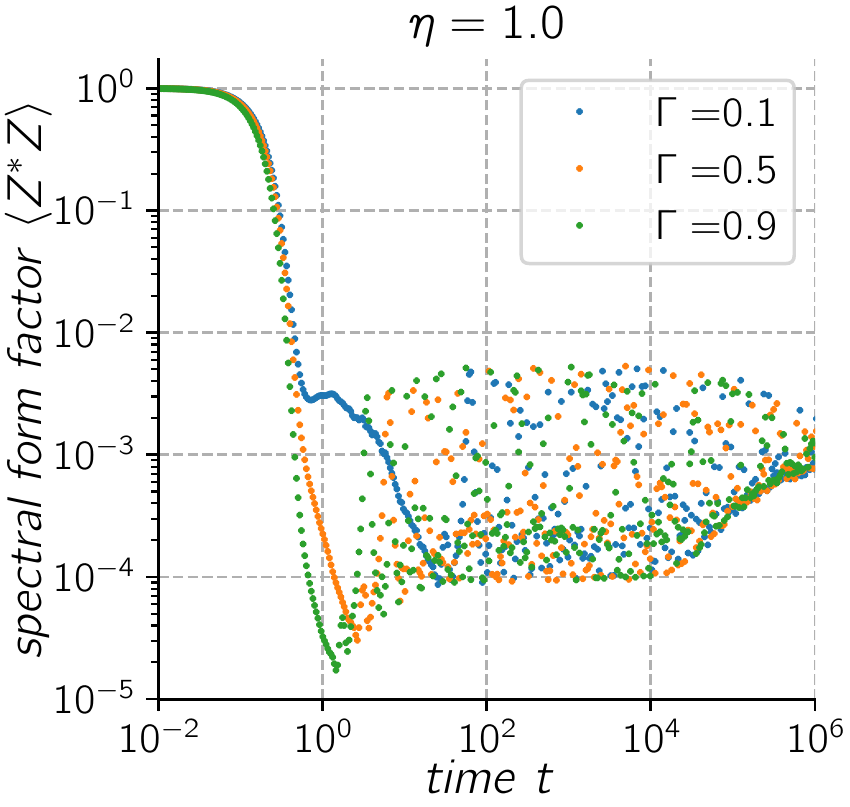}
  \includegraphics[width=0.23\textwidth]{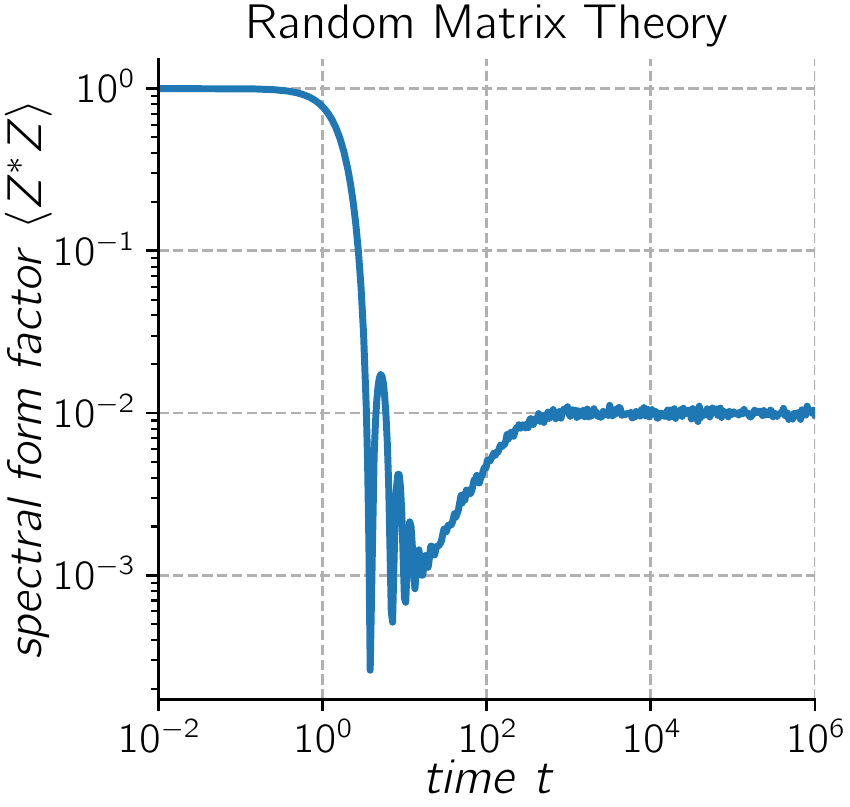}
  \caption{Spectral form factor of random dipolar spin model for 10 spins
    average over 1000 samples. We presented several constant $\eta$ cuts along
  the phase diagram which is indicated in the title. For reference, the result
from GOE random matrix theory is presented on the fourth panel.}
  \label{fig:formfactor}
\end{figure}

Spectral form factor has proven to be a useful tool for diagnosing many-body chaos in quantum systems.
For example, several works have established that $|Z(t)|^2$ shows a robust dip-ramp-plateau behavior
for the SYK model \cite{Cotler2016,Cotler2017}. Such behavior is illustrated in the
fourth panel of Fig.~\ref{fig:formfactor}
for the Gaussian Orthogonal Ensemble in random matrix
theory (with $100\times100$ matrices and thousands of random samples)
\cite{adolfo2017,adolfo2019}. The ensemble average
$\langle |Z(t)|^2\rangle$ decays rapidly at early time to reach a dip, followed by a ramp and then finally
a plateau. The result is believed to be quite generic for quantum chaotic systems.

Fig.~\ref{fig:formfactor} shows the form factor of our spin model $H_d$
computed for a few representative cuts in the phase diagram. In the Ising limit $\eta=0$, shown on the
first panel, $\langle{|Z|^2}\rangle$ decays directly to a constant value
for $\Gamma=0.1$ with no dip feature (this point belongs to the non-chaotic phase in the phase diagram). For larger
transverse field, e.g. $\Gamma=0.9$, the dip and ramp are fully developed indicating many-body chaos. At larger
anisotropy $\eta=0.5$, similar result is obtained, and
chaos is already manifest at $\Gamma=0.5$. Along the $\eta=1$ line (again
within the non-chaotic phase),
the form factor shows oscillations but no dip-ramp-plateau structure. These results are consistent with
the level spacing analysis and unambiguously identify the two regions with and
without many-body chaos in the phase diagram.

Note that the ramp connects the plateau region smoothly in the dipolar spin
model without a kink or a cusp. This is
consistent with GOE statistics as previously noted in Ref.~\cite{Cotler2016}.
In Fig.~\ref{fig:dipolar_vs_gaussian}, we study the scaling of spectral form
factor with total number of spins $N$. Different from the SYK model, smooth
connection of ramp to plateau does not change with number of spins. This means
our model is in GOE class independent of number of spins. Moreover, the early
time oscillations of form factor around the dip seems to be fading away for
larger $N$ which indicates stronger quantum chaos in $N\rightarrow\infty$
limit.
The early time oscillations before the dip seen in the RMT form factor is the
result of hard edges in the density of states (i.e. Wigner circle theorem).
Similar oscillations also takes place in SYK model since it also has hard
edges in the spectrum. Our model does not show this because of the long tails
in the density of states. Finally we observe that the dip time grows slowly
with large $N$, whereas plateau time seems to grow much faster. This is also
similar to the result reported in \cite{Cotler2016}.

\begin{figure}[h]
  \centering
  \includegraphics[width=0.3\textwidth]{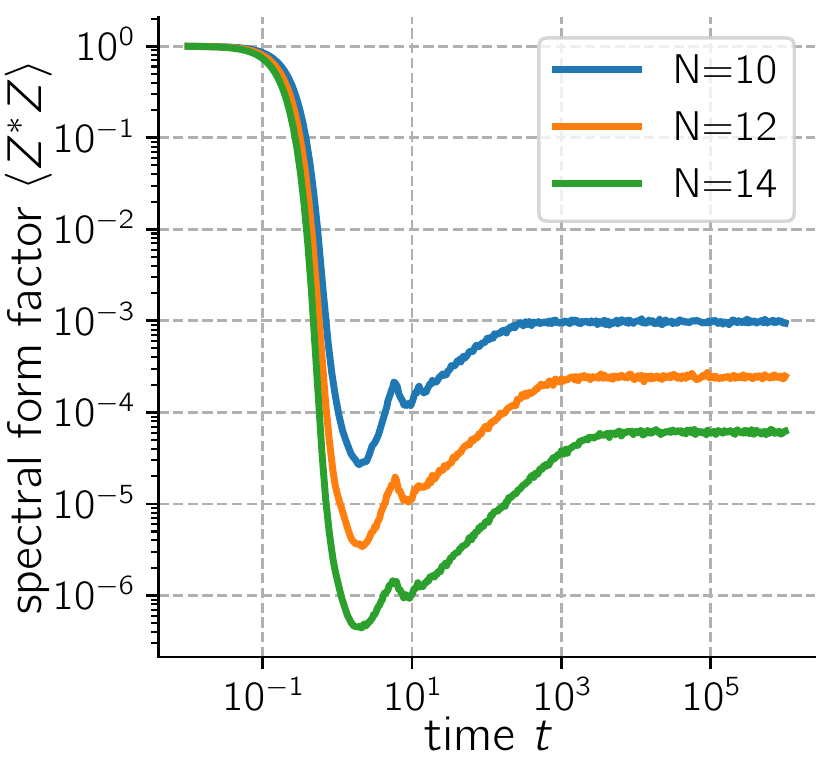}
  \caption{Scaling of the spectral form factor with the total number of spins
    in the random dipolar spin model at the quantum chaotic phase
    ($\mathcal{P}_3$ point in the phase diagram).
    Ensemble averages are calculated for 1000 random samples.}
  \label{fig:formfactorN}
\end{figure}
\section{Comparison with Sherrington-Kirkpatrick XXZ model}
\label{app:comparison} 

\begin{figure}[h]
  \centering
  \includegraphics[width=0.48\textwidth]{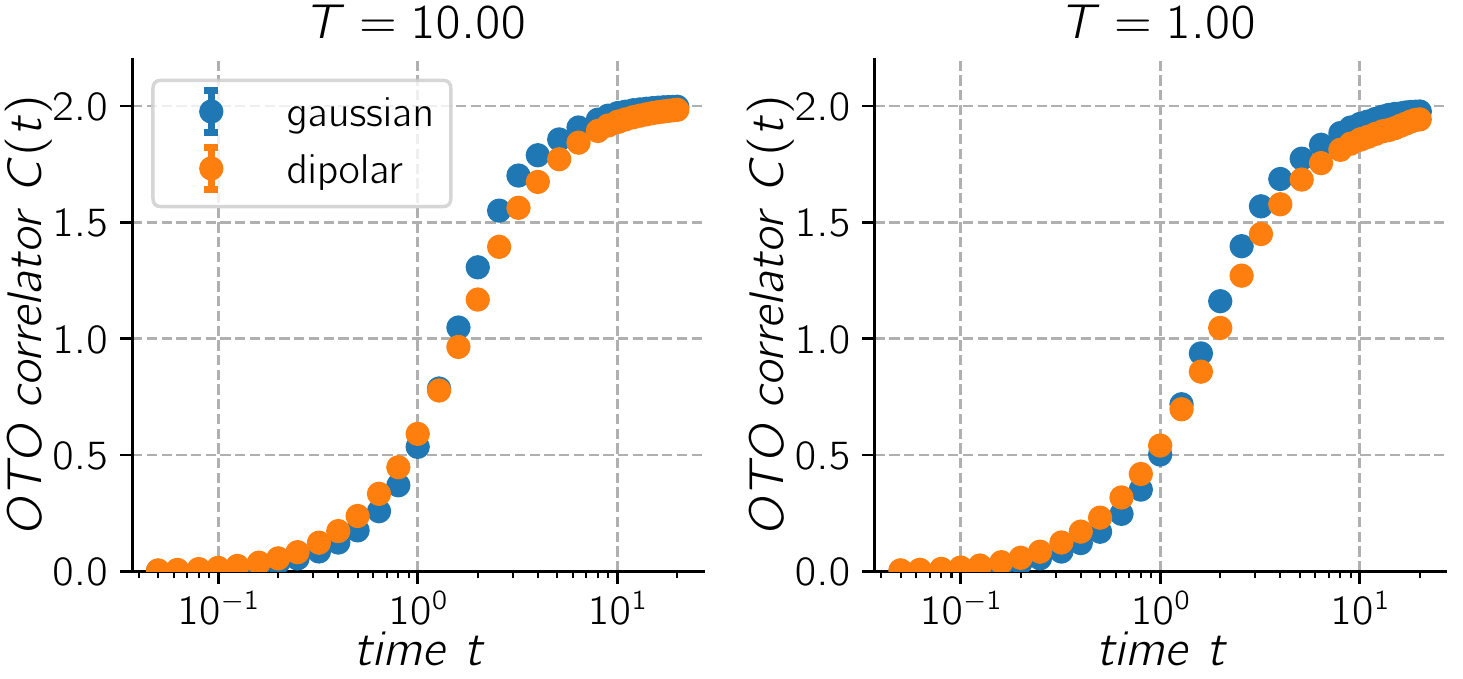}
  \includegraphics[width=0.32\textwidth]{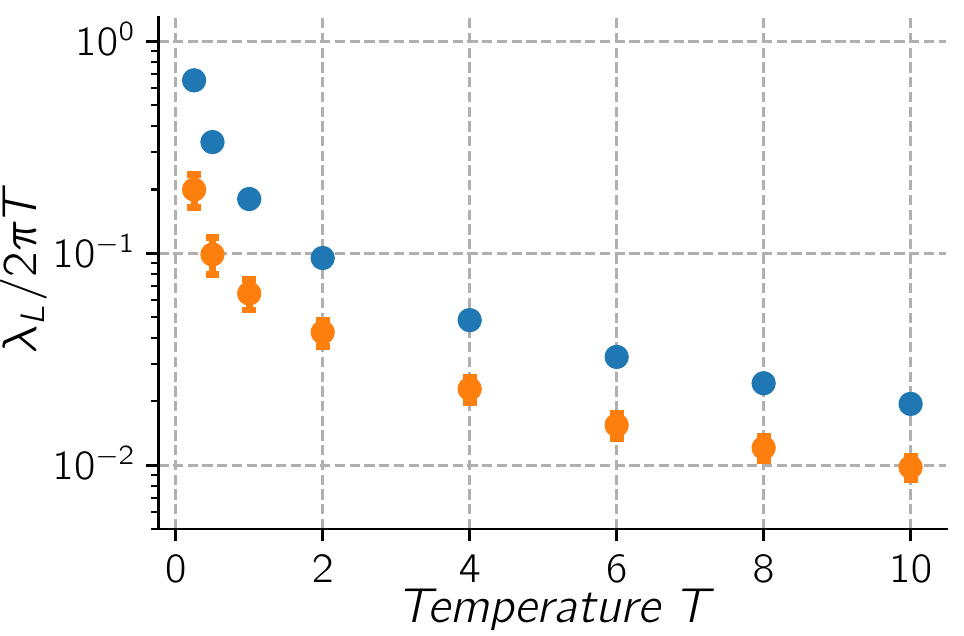}
  \caption{Comparison of OTOC of the random dipolar spins and the normally
    distributed random spins. Top left is high temperature OTOC whereas top
    right is
  the low temperature. Bottom panel shows the comparison of the scaling of Lyapunov exponent
with the temperature. For both cases, OTOC is calculated for 8 spins averaged
over 1000 random samples and $t\in[0.1,1]$ interval is used for fitting.}
  \label{fig:dipolar_vs_gaussian}
\end{figure}

In the main text, we refer to $H_d$ in Eq.~(1) with normally
distributed $J_{ij}$, with zero mean and unit variance, as the
Sherrington-Kirkpatrick XXZ model. Clearly, this model is more convenient for
theoretical investigations. In this section, we discuss our results on the
comparison of random dipolar spins with such a model with Gaussian distributed $J_{ij}$
as summarized in Fig.~\ref{fig:stat}. Comparison of the OTOC between the
Gaussian and dipolar spin models for low and high temperatures are shown in
Fig.~\ref{fig:dipolar_vs_gaussian} where the last column shows the extracted
Lyapunov exponents. Here we take
$N=8$ and $L=4$ and calculate averages over $10^3$ samples and
set $\Gamma=\eta=0.5$ which is the point
$\mathcal{P}_3$ in Fig.~1. The figure shows that dipolar spin
models exhibits a scrambling dynamics quite similar to Gaussian model for
both low and high temperatures. Extraction of Lyapunov exponent via the
same procedure explained in the main text also shows similar fast scrambling
behavior in the infrared limit. Naturally, $\lambda_L$ of the Gaussian model
seems to be slightly larger than the dipolar spin model.

\begin{figure}[h]
  \centering
  \includegraphics[width=0.32\textwidth]{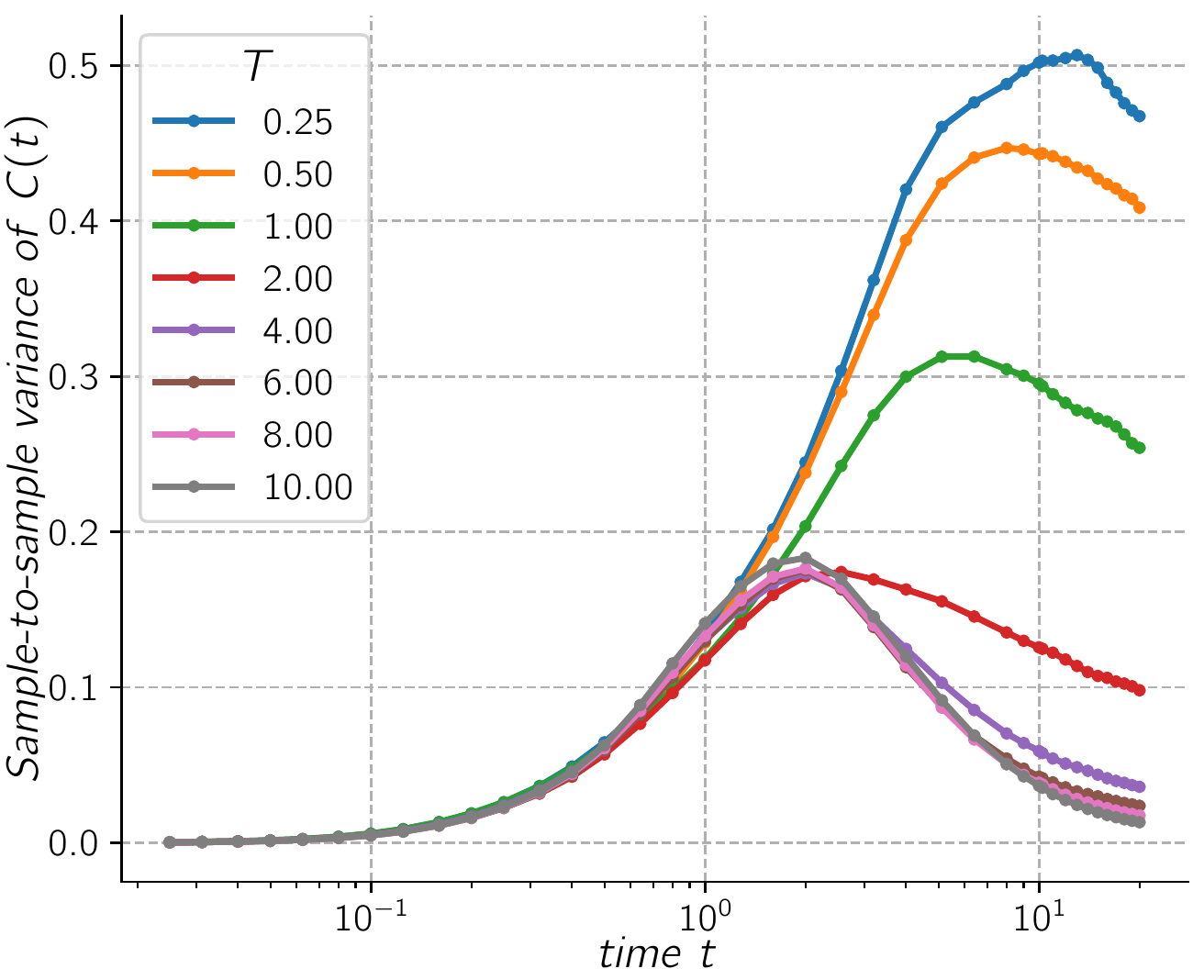}
  \caption{Sample-to-sample fluctuations of the OTOC function as a function of
    time at from low temperature $T=0.25$ to $T=10$ (Temperature increases
    from upper line to bottom line ).}
  \label{fig:variance}
\end{figure}

\section{Sample-to-sample Fluctuations of OTOC}

In this section we present our data regarding the sample-to-sample variance of
out-of-time order correlation functions $\delta
C^2(t)\equiv \dave{C^2(t)}-\dave{C(t)}^2$ where $\dave{\cdot}$ represent
disorder averaging. Our results, in Fig.~\ref{fig:variance}, show an interesting interplay between the
time scales during the scrambling and the temperature. 
It turns out that the sample-to-sample variations increases at low
temperature and is not a monotonic function of
time. For example, at high temperatures there is an intermediate time scale where the variance reaches its
maximum. 

\bibliography{refs}

\end{document}